\documentclass[aps,12pt,tightenlines,amsmath,amssymb]{revtex4}
\def\Eqref#1{Eq.~(\ref{#1})}
\newcommand{\beq}{\begin{equation}} \newcommand{\eeq}{\end{equation}}
\def\ellip{$\ldots$}
\def\gtsim{\hbox{\kern.25em\raise.5ex\hbox{$>$}\kern-.75em\lower.5ex
   \hbox{$\sim$}\kern.25em}}
\def\posonemu{\mskip 1mu}
\def\xcite#1{(\onlinecite{#1})}
\def\xeqref#1{Eq.~[\ref{#1}]}
\def\subxy{_{xy}} \def\subyx{_{yx}}
\def\fig#1{Fig.~\ref{#1}}
\def\CC{${\cal C}$} \def\CCm{{\cal C}} \def\TT{${\cal T}$} 
\def\LL{${\cal L}$}  \def\SS{${\cal S}$} \def\SSm{{\cal S}} 
\def\GG{${\cal G}$}

\def\lseqno#1{\nonumber\addtocounter{equation}1\label{#1}\leqno{[\ref{#1}]}}
\def\Jplus{J^{(+)}} 
\def\Jplust{$J^{(+)}$} 

\usepackage{epsfig}
\begin{document}

\hfill\hbox{
\small From \textit{Atti della Fondazione Giorgio Ronchi}
} 

\hfill
\hbox{\hfill \small  ~\textbf{58}, 805 (2003),\textit{with minor modifications.}}

\title{~\\ Complex systems under stochastic dynamics}

\author{L. S. Schulman}
\email{schulman@clarkson.edu}
\affiliation{Physics Department, Clarkson University, Potsdam, New York 13699-5820, USA}

\author{B. Gaveau}
\email{gaveau@ccr.jussieu.fr}
\affiliation{Laboratoire analyse et physique math\'ematique, 14 avenue F\'elix Faure, 75015 Paris, France}

\date{\today}
\begin{abstract}
\end{abstract}
\maketitle
\section{\label{sec:intro}Introduction}

A systematic study of nonequilibrium systems requires concepts and language comprehensive enough to embrace the vast collection of phenomena of interest. In our approach \xcite{master,framework,firstorder,creation} the language is that of stochastic dynamics, whose only major exclusion is quantum phenomena. Early emphasis in nonequilibrium studies was on the finding of overarching principles, analogous to entropy increase in equilibrium systems, for characterizing the steady state of a system. One candidate was the minimizing of entropy production \xcite{prigogine} which, while valuable, was nevertheless limited in its scope \xcite{kittelstatphys}. At the same time there was an acute awareness that the \textit{openness} of nonequilibrium systems led to self-organization and the formation of structure. More recently has come an appreciation that structure leads to more structure, culminating in qualitative changes in a system (emergence) often characterized by an ill-defined or perhaps multiply defined notion of \textit{complexity}.

In this article we present a stochastic dynamics approach to complex systems that is fully reductionist. No general definition of complexity will be offered; nevertheless, we intend that our considerations will help focus on the essential features of that idea.

\section{\label{sec:stochasticframework} The stochastic framework}

States of the system are modeled as points $x,y\in X$. There is an underlying stochastic process with a matrix of transition rates $R$, such that $R_{xy}$ is the probability that if the system is in state $y$ at time $t$ it reaches $x$ at time $t + \Delta t$, with $\Delta t$ usually taken to be one. We assume that $R$ is irreducible, so that its stationary state $p_0$ is strictly positive ($p_0(x)>0\ \forall\ x$). For nonequilibrium systems things can still ``happen'' in the stationary state and this feature is expressed by the existence of nonzero currents, $J_{xy}\equiv R_{xy}p_0(y) - R_{yx}p_0(x)$, describing flows of probability from one state to another. (In equilibrium one has \textit{detailed balance}, the vanishing of $J$.)

In general the only requirement for $R$ is that matrix elements are nonnegative and that columns sum to one. Irreducibility is a further convenience, and for physical, chemical and biological systems we further require that $R_{xy}>0$ if and only if $R_{yx}>0$. Extremely unlikely transitions, while unimportant for the process itself, may need to be not-quite-impossible for evaluating dissipation.

Each state $x$ is viewed as itself a coarse grain from some yet finer process and as a result will possess an intrinsic entropy, which we designate $s(x)$. This allows us to distinguish transitions which, absent external driving forces (from reservoirs), would satisfy detailed balance, i.e., $R_{xy}\exp[s(y)] = R_{yx}\exp[s(x)]$.

\section{\label{sec:inequalities} Inequalities}

Numerous general results follow from the formalism of Sec.~\ref{sec:stochasticframework}. For example, one can define the \textit{relative} entropy of two probability distributions:
\beq
S(p|q) \equiv -\sum_\alpha p_\alpha \log \frac{p_\alpha}{q_\alpha}\,.
\lseqno{eq:relativeentropy}
\eeq
This quantity increases as a system approaches its stationary state, i.e., $S(Rp|p_0)>S(p|p_0)$.

A recent result \xcite{inequalities}, bearing on complexity, is the following. Given a system as above, select a pair of states, $x$ and $y$, and change $R$ in one of the following ways:
\beq
R_{xy} \longrightarrow R_{xy}\exp(\epsilon)
\;,\qquad \hbox{or}
\lseqno{eq:changeRa}
\eeq
\beq
R_{xy} \longrightarrow R_{xy}\exp(\epsilon) \hbox{~~and~} R_{yx} \longrightarrow R_{yx}\exp(-\epsilon) \;,
\lseqno{eq:changeR}
\eeq
with diagonal matrix elements adjusted to maintain unit column sums (and $\epsilon>0$).
Then in both cases $J_{xy}$ increases. This seems intuitively obvious, but as for many results with this flavor (e.g., the GKS inequalities) the obvious is not trivial. Other apparently obvious results are not true. For example, one would expect that an increase of all matrix elements leading into a given state and decrease of all matrix elements leading out would increase the stationary probability in and total flow through that point. Nevertheless, although a numerical check for random matrices larger than 11-by-11 did not reveal any exceptions to this intuition, the obvious is in this case false. Proof of the inequality for the \textit{edge} changes given above proceeds using a representation of the stationary state in terms of spanning trees. This formula is given in \xcite{master} and apparently dates to Kirchhoff \xcite{schnakenberg}.

Our interest in inequalities arises because rate changes can occur by lowering the energy of a state or raising its entropy or by attaching a reservoir to a system. Such procedures correspond to the use of catalysts and to the introduction of negentropic resources like the sun.

\section{\label{sec:picture} The big picture}

Complexity is observed at many levels: biochemical, cellular, ecological, social, economic and others. The processes of life seem to be the most complex, but inorganic chemical reactions or meteorology can give examples as well. For our purposes systems at equilibrium have no complexity \xcite{algorithmic}. As a result we necessarily deal with \textit{open} systems, in contact with multiple reservoirs. With respect to complexity on our planet, the obvious high-temperature reservoir is the sun, a source of negentropy as well as energy, as emphasized by Schr\"odinger \xcite{schrodinger}. Other reservoirs exist, as became dramatically evident with the discovery of life in underwater thermal vents \xcite{alvin}.

Two features are thus associated with complex systems: nonzero currents ($J$) and reservoirs. The latter can be incorporated in a generalized or global detailed balance \xcite{creation}. Ordinary detailed balance gives $R_{xy}\exp[s(y)] = R_{yx}\exp[s(x)]$ (so that $p_0(x)\propto \exp(s(x))$). This extends to $R_{xy}\exp[s(y)+S(y,\eta)] = R_{yx}\exp[s(x)+S(x,\xi))]$, with $S$ the entropy of the reservoir driving the $x$-$y$ transition, and $\xi$ and $\eta$ reservoir states. Since the reservoirs are not part of the system, detailed balance is lost; nevertheless, the associated entropies are important in calculating dissipation.

In the grand scenario the sun provides energy in the form of photons, with relatively little entropy per unit energy: the surface temperature of the sun is 6000$\,$K, providing photons of more than half an electron volt, enough to drive many biochemical reactions. This energy as well as negentropy percolates through the system. Most of it either leaves earth as photons of $\sim\posonemu$300$\,$K or is locked in structured objects, such as oil (chemical structure) or sea shells (chemical and morphological structure).

A driving force of complexity is the progress of negentropy and energy through a web of exploiters \xcite{economics}. To appreciate this we elaborate on the distinction between work and heat, the essential dichotomy for the second law of thermodynamics. Work is energy in macroscopic degrees of freedom, heat is energy in microscopic degrees of freedom. One can consider work to be energy at infinite or very high temperature. Clearly these definitions depend on the distinction between macroscopic and microscopic (perhaps through a definition of coarse grains \xcite{grains}), a distinction that is more system dependent than one might expect, given the pre-eminence of the second law. Thus a leaf is able to use solar energy. Through a series of reactions it converts this to sugars and other products that do work to provide (or \textit{are}) the structure of the tree itself, as well as a storehouse of energy for reproduction. This structured (hence low entropy) energy is in turn exploited by others, from caterpillars to koala bears. (Hence some of the tree's efforts \xcite{efforts} must go into mechanisms of defense, which as for the AIDS virus can themselves become opportunistic targets.) In this picture, structure (chemical and morphological) is the tree's form of work and creatures at the next higher level use this as their resource reservoir \xcite{higher}. As a rule, inefficiency is punished. For a negentropy source there generally evolves a system (virus, fungus, cow, etc.) that can exploit it. Niches are filled to the extent possible. However, one must recognize that not all is possible. The Sahara desert or the planet Venus are both recipients of prodigious negentropy fluxes (in the form of sunlight) but lack the infrastructure to exploit it. And there are intermediate cases. Is water a primitive resource or should one include the function of sunlight in natural desalinization and treat fresh water as a product? One might include oxygen or CO$_2$ in this question, given that their planetary levels today are the result of interactions of systems driven by sunlight and other resources.

Where does complexity enter? In the competition for and in the sharing of resources, and in the filling of niches. The dung of a cow has enough structure and energy to be a resource for certain beetles. The beetle that is best at using this resource will be the one that can produce the most beetles of its own kind (provided it can defend itself against parasites and predators). Moreover, if the beetle itself is efficient the work it produces (structured objects) will provide niches for other creatures. The more efficient, the more niches. The (eventual) occupants of these niches may be harmful or helpful, but the beetle that has both efficiency and good fortune will develop symbiotic relations that will further its own cause. This same story can be told at the sub-cellular level. The use of DNA for storage and transmission of information represents the emergence of a successful biochemical mechanism from earlier mechanisms using RNA (although RNA still plays a significant role). Photosynthesis, ATP utilization, mitochondrial capture by cells, all represent situations where earlier versions did not quite optimize resource use and gave way to systems in which the degradation of negentropy was ever more gradual.

The interrelations among constituents of a system is a correlate of the niche-filling aspect and leads to interdependencies and hierarchies that qualify as complex and which may be difficult even to describe. One of the methods used by biologists, ecologists, economists and others is the \textit{flow diagram}. What happens to water? to nitrogen? Schoolchildren follow water from ocean to clouds to land to rivers to oceans. Biologists entertain more elaborate schemes \xcite{anotherbiobook,germanbiobook} 
showing the interplay of resources and actors.

Our expectation has been that the key to complexity lies in the currents, $J_{xy}$, that emerge naturally from the representation we use. These currents are directly related (by projection) to the currents and flows that appear in the descriptions recalled above. As candidates for the characterization of complexity they are far richer than the single number one would get from say, computational complexity. Besides the numerical values of the matrix elements, there is topological information in the graph associated with a given $J$. Additionally, measures of dissipation involve currents; for example, what we call Carnot dissipation \xcite{creation,schnakenberg}
can be written
\beq
{\cal D} = \sum\subxy J\subxy\log\frac{R\subxy}{R\subyx} \,.
\lseqno{eq:carnotdissipation}
\eeq
Thus in characterizing opportunities---niches---one can use the currents to spot places where negentropy is not fully exploited. As discussed earlier, $\log\left({R\subxy}/{R\subyx}\right)$ represents a change of entropy in the transitions between $y$ and $x$ (so that 
\xeqref{eq:carnotdissipation}
is analogous to electrical power, ``$IV$'').

\def\DD{${\cal D}$}

The goal, or perhaps dream, of this big picture, is a ``theorem:'' survival implies complexity. We would like to use our framework to show that the fittest systems are those that give rise to complexity. The measure of fitness however is not a simple number, and will depend on details. Consider for example the quantity \DD\ above. It would not do to take its minimization to extremes: zero \DD\ implies detailed balance. Small \DD\ can be achieved by making either $J$ small or the entropy jumps ($\log\left({R\subxy}/{R\subyx}\right)$) small. The latter is favorable. It gives opportunities for niche filling, some of which may be beneficial to the existing process. But small $J$ is (generally) not favorable. If a system is too slow a faster albeit less efficient one may overtake it in the race for survival. For the systems that do survive, it's the niche filling that induces complexity. As a theorem, our assertion could only be probabilistic, since as we are all aware a few nuclear bombs could radically ``simplify'' our planet. Less evident dangers may also exist; an imaginative and fortunately fictional possibility is Vonnegut's \textit{ice-nine} \xcite{vonnegut}.
A second problem involves intrinsic limitations. For example, given the laws of physics and the resources of our planet, it may be that the system long ago attained a maximum of complexity; subsequent changes have been fluctuations around that level. We may be no more complex than dinosaurs, merely different. The contemporary emergence of life-like functions on silicon (etc.) may represent an extension of the potential for complexity at the microscopic level. Lacking a quantitative handle, it's not possible to be definitive on this point, even if one occasionally feels that the Windows operating system represents a new and malevolent life form.

In our formal development two perspectives have been used, depending on whether or not one wishes to include evolution. One approach is to take a given system, such as the biochemistry of glycolysis or the interaction of koala bears and eucalyptus trees, and analyze its fluxes and complexity using a fixed set of states and transition probabilities. Alternatively, one can allow changes in the rules---a meta-stochastic process in which $R$ is itself a random variable---to watch the emergence of complexity as negentropy resources are exploited. The latter is the more ambitious approach and involves an extension of our earlier formalism. Nevertheless, this approach can be looked upon as an approximation to a \textit{much} larger version of the original formalism in which the space ``$X$'' includes the dynamics of smaller units (atoms?) and gets closer to the basic laws of physics. For this larger space one is interested not in the stationary state (which may never occur), but rather in the long-lived metastable states, which the $R$s of the meta-process have as their stationary states.

\section{\label{sec:models} Models}

To carry out the program described above we start with simple systems. An example, drawn from ecology \xcite{gotelli}
(and applied more widely \xcite{feigenbaum}%
), is the discrete-time logistic equation
\beq
N(t+1)=N(t)+rN(t)\left(1-\frac {N(t)}{K}\right) \,,
\lseqno{eq:logistic}
\eeq
where $N(t)$ is the population of say, rabbits, at time $t$, $r$ is their reproduction rate and $K$ the ``carrying capacity,'' the number of rabbits supportable in the steady state in the area of interest. As in the more familiar parameterization used in physics ($x'=\lambda x(1-x)$), with increasing $r$ this system goes from a stable state with $N=K$ (for $0<r\leq 2$) to a period-2 stationary state, to period-4 and ultimately to chaos and beyond \xcite{beyond}.
This can be recast as a stochastic dynamics by coarse graining, putting populations into bins of size $N_{\hbox{\small max}}/G$, with $N_{\hbox{\small max}}$ the maximum for given $r$ and $K$, and $G$ the number of grains \xcite{russell}.
The transition probability from grain to grain is proportional to the number of points in the image grain under the mapping \xeqref{eq:logistic}.
The stochastic dynamics shows no currents (as a function of increasing $r$) until the second period-doubling bifurcation, subsequent to which there are increasing currents as well as loop structure (in $J$) that reflect the stationary orbits of the system. Approaching chaos, the loop structure shows as much of the true structure as the coarse graining permits. In parallel, an overall increase in current with $r$ is found. This model is obviously a warm up---for us as for ecologists. Additional actors should be introduced, perhaps grass, lynxes, sunlight, water. Sunlight is the ultimate resource, both for the growth of the grass and for the recycling of the water (although without a basic water reservoir the sunlight would not help), so that if one used a ``$K$'' (as in \xeqref{eq:logistic}) for the grass, it would no longer be a parameter, but a function of the state of the other systems---leading to rules for coarse grained transition probabilities that reflected those dependencies. Similarly rabbits play the role of a reservoir for the lynxes. At this stage, study of this system can help clarify the role of currents in the description of complexity. Whether it can contribute to ecology is less clear. The mantra is that understanding complexity helps understand \textit{everything}; nevertheless one would like more immediate benefits. Perhaps the systematic definition of coarse grains \xcite{grains}
can lead to the identification of critical factors or critical emergent quantities in such systems (as such quantities were objectively derived in Ref.~\xcite{grains}).

\begin{figure}
\centerline{\epsfig{file=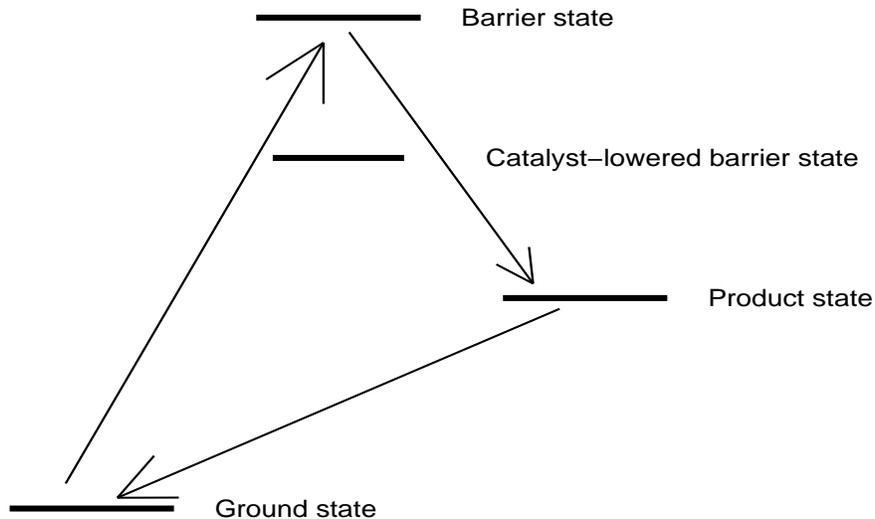,height=.3\textheight,width=.5\textheight}}
\caption{Schematic diagram of the context and action of a catalyst. The arrows represent the flow of current when the G$\to$B process occurs at higher temperature than the B$\to$P process. For the matrix $R$, non-zero matrix elements exist in both directions.}
\label{fig:catalyst}
\end{figure}

The processes of life, studied at the biochemical level, should also reveal the sources of complexity. Consider a model of catalysis, perhaps through the agency of an enzyme that lowers energetic or structural (entropic) barriers. \fig{fig:catalyst}
is a schematic diagram of a typical situation: a reaction is impeded by a barrier; in the presence of a catalyst the barrier is lowered. For example, an enzyme that presents extended reactants to one another in such a way that active regions come in contact, can be considered to have lowered an entropic barrier. The situation in \fig{fig:catalyst} can also be treated as a Carnot cycle, taking the process G$\to$B as occurring at temperature $T_1$ and B$\to$P at temperature $T_2$, with $T_1>T_2$. In the course of P$\to$G energy is released. This is the crucial step (the potential ``profit''), and by the usual estimates of Carnot efficiency one could draw work proportional to $1-T_2/T_1$ from this transition. One form this work could take is the creation of structure in the form of the creation of additional enzymes, either for this very process or for others. The portion of energy not so used (or otherwise exploited as work) is wasted and will show up as dissipation through the extreme smallness of the transition probability from~G directly to~P. Within this context one can also recall the concept of \textit{catalytic power} \xcite{timebook},      
and relate it, as expected, to rate variation and entropy shifts (specifically when a barrier is lowered, there is an increase in rate, usually related to the increase in entropy \xcite{inequalities}).

A process that could directly produce its own catalysts could take over the world! Physically the common situation is that the catalyst produced by one reaction helps in another. If a sequence of such directed links closes on itself, the consequences are twofold: the set of reactions is speeded up and the interdependency creates (or \textit{is}) complexity.

Another biochemical source of complexity is the \textit{ladder}. This mundane concept plays a role when available energy packages are not big enough for the task at hand. Suppose the ``task'' is to overcome a barrier of height $E_B$, but the energy-source temperature is only $kT\sim E_B/2$. The waiting time for this process is on the scale of $\exp(E_B/kT)$. With an appropriate ladder this can be reduced to $\sim2\exp(E_B/2kT)$. Here is the principle: a ``ladder'' consists of a pair of levels, call them $a$ and $b$, interposed between G and B, whose role is to help the system ``climb'' from G to B. Take $E_a\sim(E_B-E_b)\gtsim E_B/2$ ($E_G =0$ and the connections are \hbox{G$\,\leftrightarrow a\leftrightarrow b\leftrightarrow\,$B}, an extension of \fig{fig:catalyst}). In the temperature-$T$ environment the system (relatively) easily climbs to the $a$ state. From $a$ it can fall back to G or it can drop to $b$. The latter drop takes place at a much lower temperature. Such a situation is commonly encountered when a molecule is excited (to $a$) by an external source (relatively high energy photons, say), but once in $a$ drops to $b$ through the emission of phonons---relaxation of the molecule, with a cool solid or solvent carrying away the phonons. Once in the state $b$ it is again well-coupled to the external photons and makes its second jump, to B. This system can easily be modeled as a stochastic process, allowing the system, once in the barrier state, to drop to the product state and finally back to the ground state, as in \fig{fig:catalyst} (but now with $a$ and $b$ levels interposed between the ground and barrier states). If the temperature in the $a\to b$ drop is about a tenth of the external source temperature (for reasonable energy-level values) the current can increase by a factor 4. Intuitively this is easy to understand: a ladder is useful if you do not slip or bounce on the rungs. An example of the use of the ladder principle is the breakup of glucose by a series of ATP-induced processes. (The fragmented glucose is used to convert ADP to yet more ATP, but this is part of the ``work'' phase of this cycle.) A second biochemical example is photosynthesis \xcite{germanbiobook}.

Another model that we have studied is ecological: koala bears and eucalyptus trees. The trees benefit from external sources (sunlight, water), while the bears get all sustenance from the trees. Here we looked at the \textit{meta}-stochastic process. Starting with little coupling between the two systems, random ``evolution'' was allowed. Matrix elements in the combined system were allowed to increase or decrease and the satisfaction level (number of individuals, biomass) of each species tested as to whether a given ``mutation'' would survive. Feedback from bear to tree was also allowed, for example, the bear might serve as a source of fertilizer for nutrients needed by the tree. Important features of the real world were absent in these simulations; nevertheless, it was found \xcite{koala}
that as satisfaction levels increased, so did currents.

\section{\label{sec:efficiency} Efficiency and complexity}

As seen in Secs.\ \ref{sec:picture} and \ref{sec:models},
lowering dissipation can lead to increased participation by other actors: beetles to take care of waste products, arbitrageurs, coupled creators of mutually beneficent enzymes. Clearly these lead to an increase in currents, and according to our intuitions, to complexity. How to measure this notion?

First one must consider whether currents ought to be sufficient for the characterization. From $J$ alone it is not possible to recover $R$ (nor \DD, [\ref{eq:carnotdissipation}]).
Nevertheless, we focus on $J$, bearing in mind the utility of rate equations in chemistry, equations derivable from the currents alone, ignorant of more detailed mechanisms of a reaction.

Comparing two $J$s and their associated graphs, if one graph is a subset of the other with proportional currents, then one would consider the larger graph the more complex; a \textit{partial} order is clearly possible. To go beyond this, perhaps to a topological definition or perhaps to a single number, further considerations should be invoked.

\begin{figure}
\centerline{\epsfig{file=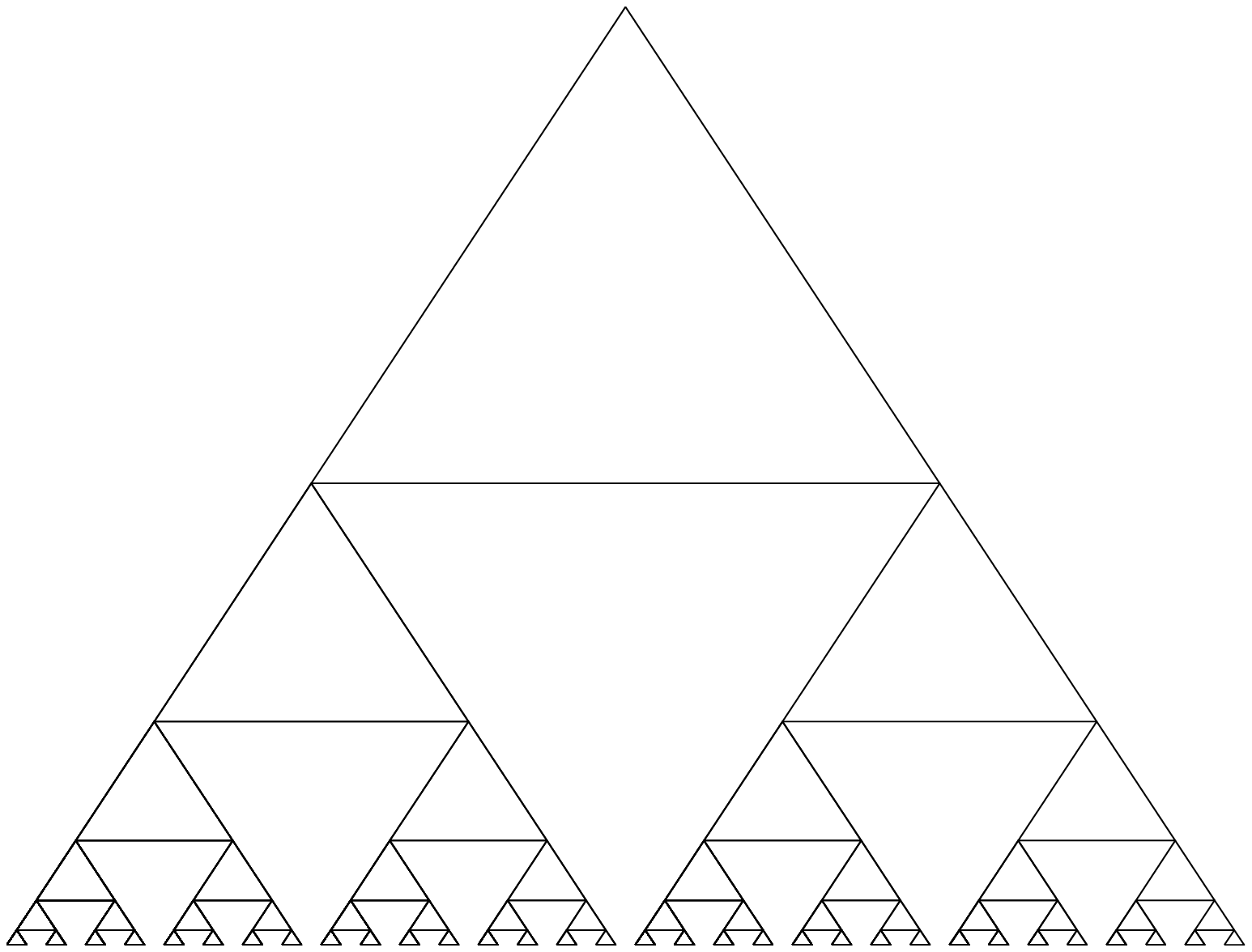,height=.3\textheight,width=.3\textheight}
\epsfig{file=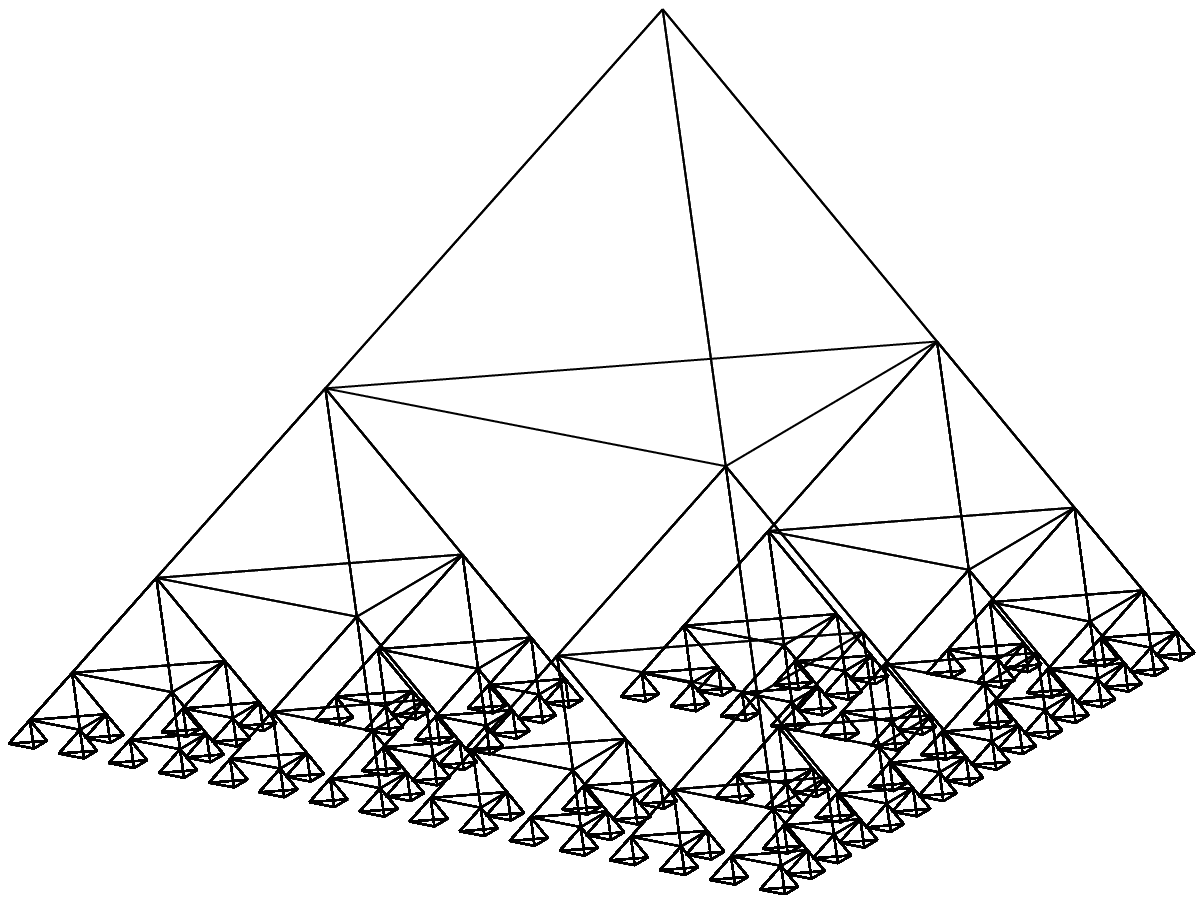,height=.3\textheight,width=.3\textheight}}
\caption{Self-similar graphs.}
\label{fig:selfsimilar}
\end{figure}

Recalling the ease of definition of \textit{information} based on simple axioms \xcite{katz}
(easier in retrospect than prospect), we consider what \textit{complexity}, to be called \CC, should satisfy. First, extensivity would be inappropriate: one grass seed will (eventually) create the same lawn as two identical seeds. So there needs to be a combining rule that makes two cats more complex than one cat but less complex than one cat and one grasshopper. This suggests that while entropy (or information) is the logarithm of a large number, \CC\ ought to be log-log. A second requirement or axiom has to do with coarse graining. In discussing this, we make the point that complexity is the complexity of a \textit{description}. The ecologist who wants to compare the complexity of the rain forest with that of a wheat farm will not have reason to include internal cellular processes (although genetically modified crops are changing this). So one expects that \textit{coarse graining}, smearing out the most detailed information, \textit{should reduce \CC\ by an order unity fraction of itself}. A precise version of this requirement can be formulated for the case where $\Jplus$ has self-similar structure. Thus in \fig{fig:selfsimilar}
eliminating (all 3 sides of) the topmost triangle of the 2-dimensional illustration would create two copies of a simpler object, which should have the complexity of a single exemplar of that simpler object (perhaps with a log-log correction, just as entropy is extensive up to surface effects). However, coarse graining (say) the bottom-most row creates the same object, presumably giving the same value for complexity. This argument assumes the values of the currents are similarly scaled.

This leads us to discuss overall scales. For the systems considered, $X$ will usually be a product space. There is some number, $M$, of individual actors (atoms, cells, people, species, \ellip); these have a variety of internal states. For simplicity we take these all of the same finite cardinality, say $K$. This gives $X$ (which is a product of the ``actor'' spaces) dimension $N\equiv K^M$. Let $\Jplus\subxy=(J\subxy+|J\subxy|)/2$, the matrix of positive currents. The maximum number of non-zero components of $\Jplus$ is $N(N-1)/2$, but they are not independent: the sum of \textit{in}- and \textit{out}-currents at each vertex must be equal, yielding $N - 1$ constraints (since one equality is redundant). The graph of $\Jplus$ will thus have $E$\ ($\leq N(N-1)/2$) directed edges with $E-N+1$ independent quantities.

A number of possibilities for defining \CC\ have been considered, none quite right (as far as we can tell) but each embodying to some extent the ideas discussed above.

It is often informative to write \Jplust\ in terms of its loops. We have considered loop decompositions of \Jplust\ in a number of ways. Numerically a decomposition can be found by selecting the edge with the smallest current and completing a non-self-crossing loop. Remove this loop and continue. Other decompositions are possible using bases generated by trees~\xcite{schnakenberg}.
A decomposition of \Jplust\ has the general form $\Jplus=\sum_\alpha u_\alpha J^{(+\alpha)}$, where $J^{(+\alpha)}$ is the current associated with the permutation operator around the loop $\alpha$. Generally, for graphs beyond the simplest, the expansion is far from unique and in fact the set of coefficients $\{u_\alpha\}$ form a simplex of high dimension. Each $u_\alpha$ has minimum value 0 and a maximum that depends on \Jplust. A definition of \CC\ could involve properties of this simplex, for example the log-log of its dimension or (appropriate logarithms of) sums over functions of the $u_\alpha$'s.

Another, and computationally preferable method, works directly from the components of \Jplust. Let \LL\ be the set of all possible loops and \TT\ the set of all distinct connected subgraphs of the graph, \GG, of \Jplust. Consider an individual graph $\alpha\in\SSm$, where \SS\ is either \LL\ or \TT. Go through \GG\ and find all exemplars of $\alpha$. With each of these associate a current or current matrix, namely the maximum current (matrix) on this graph consistent with the original \Jplust. (If $\alpha$ is a cycle this will be the minimum value of \Jplust\ along the cycle. For other graphs the determination may be more complicated.) A candidate for \CC\ is the following
\beq
\CCm = -\log \left[ - \sum_{{\cal S}}\log \sum_{\alpha\in {\cal S}}\prod_k j^{(k)}_\alpha\right] \,,
\lseqno{eq:complexcurrents}
\eeq
where $j^{(k)}_\alpha$ is the current along edge $k$ of the graph or loop $\alpha$. A form that seems to work a little better looks only at the loops and drops the product in \Eqref{eq:complexcurrents}:
\beq
\CCm = -\log \left[ - \sum_{{\cal L}}\log \sum_{\alpha\in {\cal L}} j_\alpha\right] \,,
\lseqno{eq:complexcurrentsloopsonly}
\eeq
with $j_\alpha$ the maximal current through the loop $\alpha$. Both definitions have the property that two identical grass seeds increase \CC\ only slightly, while an apple and a grass seed will give rise to many different sorts of structures and increase \CC\ more markedly. Non-interacting cats will be like non-interacting grass seeds, but slightly interacting cats (giving rise to many, many new structures) will nevertheless not increase \CC\ disproportionately, because of the presence of the actual current values in the products.

\section{\label{sec:discussion}Discussion}

In this picture complexity arises in the exploitation of resources. Exploitation includes secondary resource creation, competition, sharing and degradation. Because our focus has been on natural systems, the basic rules are the first and second laws of thermodynamics as well as other relevant constraints from physics and chemistry. An economic or sociological model would identify other resources and constraints.

Efficiency, meaning the minimizing of dissipation, plays an ambiguous role. An efficient process wields a two-edged sword. Like any non-equilibrium system or subsystem it operates between reservoirs, perhaps the primary reservoirs of sun and outer space or more frequently between intermediate constructs, like food and feces or ATP and \hbox{ADP}. Consider the fruit of a tree. Its structure and content are part of the \textit{work} produced in the Carnot cycle in which the tree exploits its reservoirs. The efficient tree will produce many, rich fruits. These will be reservoirs for other creatures. Some of these may be pathogens that gain a foothold in the fruit and go on to destroy the tree. Others may eat the fruit and in their feces spread the seeds, to the tree's ultimate advantage. The system lucky enough to have productive symbiotes will survive. Being efficient in this case ensures more tries at the dice table. But there is a downside to efficiency, at least as defined through 
\xeqref{eq:carnotdissipation}.
That expression also contains the currents, which is to say that efficiency is gained by slowing down. (In the textbook Carnot cycle the processes are "quasi-stationary," ensuring reversibility). In the world of competing and self-reproducing processes (whether animals or enzymes) speed does matter: the quicker process will grab the resources even if its use of them is less efficient. There is nothing wrong with the ambiguous role played by efficiency: compromise is the essence of good engineering (and \textit{a fortiori} of politics and economics). Nevertheless, these considerations do imply that there is no general principal to the effect that efficiency implies survival. What does emerge though is that the role of efficiency in survival enhances complexity. This is because the virtue of efficiency is felt when another player is able to make use of the work product in a way that enhances the survival of the producer. Such interdependencies are the essence of complexity.

\def\tot{$\to$} 

The general discussion of the previous paragraph can be realized in greater detail in one of the models discussed above. \fig{fig:catalyst} shows a ground state (G), barrier (B) and product (P). Recall that the process G\tot B occurs at higher temperature than \hbox{B\tot P}. A higher energy value for P is better, in that it allows more work, here proportional to the P-G energy difference. On the other hand, raising P slows the process, allowing more backtracking (the reverse reaction, P\tot B, can dominate, depending on the energies and temperatures). The exploitation of the work can take various forms. For example, the work produced could be expressed as a part of the product, P, which could be a catalyst. If it is a catalyst for the 
G\tot B\tot P reaction the process will take off, limited only by available resources, perhaps a supply of the constituents of G or by the reservoirs being less than ideal and having finite capacity. If the product is a catalyst for some other reaction, that other reaction could itself be producing a catalyst. If a network of such catalysts closes one obtains a successful ``creature,'' perhaps meriting the name ``life.''

\def\remark#1{\medskip\noindent{\textit{Remark:~}} #1}

\remark {In a complex system it is often the case that the utility of a structure or process is expressed at the next higher level of organization relative to the process you're studying.}

\remark {It seems that randomness is remarkably adept at exploration. Although they cannot literally be true, to a good approximation the following maxims appear to be obeyed:\\
Everything is possible becomes actual---if you can wait long enough.\\
Things \textit{must} happen because they \textit{can} happen (same temporal limitation).}

\medskip

Finally, we struggled with ideas for a definition of complexity. We expect that its definition should be richer than that of algorithmic complexity, and should express the level of interconnectedness and interdependencies of a system, not just the instruction set for creating the system---\textit{the amount of effort it takes to use those instructions must addressed as well}. Unlike entropy and the related concept of information, complexity is not extensive, nor is it entirely intensive. Twenty cats will be more complex than one. The entropy per unit volume of 20 blocks of ice will be the same as that of one block, up to surface effects that will be entirely negligible for a macroscopic piece of ice. The twenty cats however will develop relations among themselves that qualify for an additional level description and interdependence.

For this reason our tendency has been to say that as entropy is log, so complexity is log-log. Unfortunately we don't know \textit{what} it is the log-log of. What is clear though is that complexity is the complexity of a specific \textit{description}, which is of course dependent on the technology and subjective capabilities of the observer. In any case, we presented candidates for a quantitative notion of this ``\CC''; for each of them we found ways in which it captured desired features and ways in which it fell short.

\begin{acknowledgments}
We thank Christine Bertrand, Charles Rockland and Leonard J. Schulman for helpful discussions. This work was supported in part by the United States National Science Foundation Grant PHY 00 99471.
\end{acknowledgments}

\begin{thebibliography}{27}
\bibitem[{alg()}]{algorithmic}
\bibinfo{note}{Note that our questions are different from those that arise in
  \textit{algorithmic complexity} \xcite{chaitin,lloyd}, although there may
  ultimately be overlap with those ideas.}
\bibitem[{eff()}]{efforts}
\bibinfo{note}{These ``efforts'' are themselves the work product, as discussed
  in the next sentence.}
\bibitem[{hig()}]{higher}
\bibinfo{note}{The term ``higher'' is used here in a conventional sense, but
  the food chain respects no nobility. Viruses kill as well as tigers. The
  prion mindlessly preys upon, or takes as a resource, objects possessing
  ``higher'' levels of organization.}
\end{thebibliography}

\end{document}